# Quantum Beating in Ring Conductance: Observation of Spin Chiral States and Berry's Phase

M. J. Yang[*], C. H. Yang[†], K. A. Cheng[†], and Y. B. Lyanda-Geller[*,‡]

[*]Naval Research Laboratory, Washington, DC 20375

[†]Dept. of Electrical and Computer Engineering, Univ. of Maryland, College Park, MD 20742

[‡]Beckman Institute and Department of Physics, Univ. of Illinois, Urbana, IL 61801

*Abstract*

Using singly connected rings with a collimating contact to the current leads, we have observed the spin quantum beating in the Aharonov-Bohm conductance oscillations. We demonstrate that the beating is the result of the superposition of two independent interference patterns associated with two orthogonal spin chiral states arising from intrinsic spin-orbit interactions. Our work provides the conclusive evidence of the spin Berry's phase in the conductance of quantum rings.

When a quantum system evolves adiabatically through a cyclic variation of its parameters, the physical state acquires a memory of its motion in the form of geometric and dynamical phases in the wavefunction[1,2]. In contrast to the dynamical phase that records the cycle duration, the geometric phase depends only on the path traced out in the parameter space. Since the discovery by Berry, geometric phases have been demonstrated experimentally with, e.g., polarization rotation of photons, neutron spin rotation and nuclear magnetic resonance.[2] Extensive theoretical and experimental work have shown that Berry's phase is a general phenomenon in various fields ranging from elementary particles to chemical and condensed matter physics. Recently, due to the potential applications for quantum computing[3] and spintronics,[4] electron spin in semiconductor quantum devices has become the center of attention. One of the critical issues for quantum computation is the ability to sustain electron spin coherence. The observation of spin interference phenomena such as spin quantum beating and manifestations of the spin Berry's phase in ring conductance is a crucial milestone on the way to spin-coherent quantum circuits. To date, the spin quantum beating in ring conductance has never been observed, and therefore the results of the pioneering studies of Berry's phase in semiconductors[5,6,7] has been questionable. The challenges to its identification have been in the complexity of data in devices with multiple electronic modes and the requirement of adiabatic transport. Here we report the first observation of the spin quantum beating in the Aharonov-Bohm (AB) conductance oscillations[8,9] and demonstrate that the beating is associated with spin chiral states arising from intrinsic spin-orbit (SO) interactions. Our work provides the first conclusive evidence of the spin Berry's phase in the conductance of quantum rings.

Mesoscopic semiconductor devices[10] exhibit long mean free paths and long phase coherent lengths and provide experimental systems in which quantum phases can be observed.

Transport studies in mesoscopic rings have demonstrated conductance oscillations due to the AB interference, where charged particles acquire an additional phase $2\pi\Phi/\Phi_0$ after completing a closed circuit. Here $\Phi_0 = h/e$ is the magnetic flux quantum, and $\Phi$ is the total flux enclosed by the ring. The subject of Berry's phase in mesoscopic systems was first introduced and explored[11] in the context of conducting rings under magnetic textures. It has also been proposed[12] that conducting rings provide a physical setting in which the spin Berry's phase can be observed as a result of intrinsic SO interactions. If the SO interaction originates from the asymmetry of the confinement potential,[13] the electrons in a ring would experience a radial built-in Zeeman-like magnetic field ($B_{in}$ in Fig. 1). The $B_{in}$, whose amplitude is proportional to the electron longitudinal momentum, results in spin splitting and leads to the formation of two chiral spin states. When electrons adiabatically encircle a ring under an external magnetic field $B_{ext}$, electron spin, influenced by the total effective magnetic field ($\vec{B}_{eff} = \vec{B}_{ext} + \vec{B}_{in}$) subtends a cone-shaped trajectory in parameter space, as illustrated in Fig. 1 (a). Here $\vec{B}_{ext} = B_{ext} \cdot \hat{z}$ is along the sample growth direction. Because the geometric phase is given by half the solid angle subtended by $\vec{B}_{eff}$ for particles with spin 1/2, the electron wavefunction acquires a spin Berry phase of $\pi(1-\cos\theta)$, where $\theta = tan^{-1}(B_{in}/B_{ext})$.

Manifestations of Berry's phase in mesoscopic rings have been intensively studied theoretically[11,12,14,15,16] in persistent currents, tunnelling, weak localization and electron-electron interactions effects in conductivity. Experimental attempts to probe Berry's phase in doubly connected rings[5,6,7], concentrated on its signatures in Fourier spectra of AB oscillations in the presence of SO interactions. However, the absence of a beating pattern in AB oscillations makes the identification of Berry's phase difficult. Moreover, in order to simplify the analysis, two important issues have been largely ignored, namely, achieving the single mode transport and the presence of spin rotations in the contacts. Even with the current advance of nanotechnology, it is still a challenge to fabricate single mode nanowires. In the doubly connected rings previously used, all transverse modes in the lead are likely to enter the ring, thereby making the conductance oscillation pattern difficult to analyse. The other obstacle in the doubly connected configuration is the complication of electron transmission at the contacts, which was simplified in Ref. [12] in order to elucidate the geometric phase due to electron motion in a ring and has been ignored since then.

The transmission of electrons in spin chiral states through contacts has profound influence on the manifestation of Berry's phase[17] and is rather complicated in a doubly-connected ring. Figure 1 (b) shows the critical influence of spin rotation at the contacts, where we consider the ring and the leads as one coherent system and assume adiabatic passage of electrons. When electrons enter the system from the lead and travel along the ring, the corresponding trajectory in parameter space encloses no area, i.e., electrons do not acquire any geometric phase after a single passage, as shown in Fig. 1(b). In other words, the geometric phase accrued in the ring is cancelled by the phase acquired due to spin rotations in the contacts. However, transmission through contacts can also have a sudden, non-adiabatic character, with no spin rotation. In this case, reflection from the contact can play a significant role, and electron transmission has to be treated as resonant tunnelling. Berry's phase then can manifest itself in resonant tunnelling spectra.[12] Consequently, in any realistic study of quantum phases in doubly-connected rings, one needs to take into account spin rotations in the contacts, and the degree of non-adiabaticity of passage.

To overcome these challenges, we have designed a singly connected ring with a collimating contact[9] to the current leads, as shown in Fig. 1(a). In this configuration, it is possible to let only one transverse mode with a small longitudinal momentum enter the ring through the contact. Electrons in other modes, which have larger longitudinal momentum, prefer to propagate in the straight current lead and bypass the ring. As a result of this momentum filtering by the contact, the spin quantum beating pattern in the conductance is determined solely by a single transverse mode. Note that in the ring region, the high momentum modes, although decoupled to the current leads, are still present to screen out the potential fluctuations due to intrinsic impurities. As a result, this system possesses a unique property: the interference comes solely from a single transverse mode that has a long phase coherence length.

For the electron mode that enters the ring, the full spin phase $\varphi_\pm$ acquired by the electron wavefunction as the result of a single adiabatic passage of a ring can be expressed as

$$\frac{\varphi_\pm}{2\pi} = k_0 R \pm \frac{\sqrt{\omega_Z^2 + \omega_{SO}^2}}{\Omega} \pm \frac{1}{2}\left(1 - \frac{\omega_Z}{\sqrt{\omega_Z^2 + \omega_{SO}^2}}\right) + \frac{\varphi_{AB}}{2\pi} \qquad (1).$$

Here +/- indicates the two chiral states, and $k_0$, $R$, $\hbar\omega_Z = \frac{1}{2}g^*\mu B_{ext}$, $\hbar\omega_{SO} = \alpha k_0$, $\Omega = \hbar k_0 / m^* R$, and $\frac{\varphi_{AB}}{2\pi} = \frac{\Phi}{\Phi_0} = \frac{\pi R^2 B_{ext}}{h/e}$ are the wave vector of spin-degenerate electrons, radius of the ring, Zeeman energy, SO energy, the semiclassical frequency of electron rotation around the ring and the AB phase, respectively, with $g^*$ the electron effective $g$-factor, $\mu$ the Bohr magneton, α the SO constant and $m^*$ the electron effective mass. The corresponding SO magnetic field is $B_{in} = \frac{2\alpha k_0}{g^*\mu}$. The first two terms in Eq. (1) represent the dynamical phase, while the third term describes the spin Berry's phase. In the absence of $B_{ext}$, the phase difference between the two chiral states is $\Delta\varphi = \varphi_+ - \varphi_- = 2\pi[(2\omega_{SO}/W)+1]$. In contrast to the monotonic increase of $\Delta\varphi$ due to the spin dynamical phase on the increasing $B_{ext}$, $\Delta\varphi$ due to the spin Berry phase is decreased monotonically by $2\pi\cos\theta$. Note that the adiabaticity requires $\omega_{SO} > W$ (or $\frac{\alpha m^* R}{\hbar^2} > 1$), i.e., a large α so that the electron spin precesses a few times within a cycle.

When the spin dephasing length is larger than the perimeter of the ring, the total conductance contributed by the two chiral spin states can then be expressed as

$$G = \frac{e^2}{h}[2 - \sum_{\pm} \frac{(1-e^{-2\delta})}{2+e^{-2\delta}-2\sqrt{2}e^{-\delta}\cos(\varphi_{\pm})}] \qquad (2).$$

Here we assume for simplicity that electrons enter the ring with an amplitude $h = \sqrt{2}/2$ without reflection.[18] Then the interference signal arises in the presence of electron dephasing in the ring, modeled by transfer coefficient $\exp(-\delta) = \exp(-2\pi R/L_f)$, where $L_f$ is the phase coherence length.

In this work, we use AlSb/InAs/AlSb single quantum wells grown by molecular beam epitaxy with a well width of 17nm. The measured two-dimensional density and mobility are 4.9 × $10^{15}$ m$^{-2}$ and 20 m$^2$/Vs. The corresponding Fermi energy, elastic mean free path and thermal length at 4K are 43 meV, 2.3 μm and 4.2μm respectively. We employed a newly developed nanofabrication technique[19] to process several rings with radii of 150nm, 250nm, 350nm, and 500nm. The right inset in Fig. 2 (a) displays an atomic force microscope (AFM) image of a 350nm ring. The lithographic width of the wire is 95nm, and the estimated conducting channel width is 70nm.[20] The magnetoresistance of a 500nm ring is shown in the left inset of Fig. 2 (a). In addition to the reproducible aperiodic fluctuations due to the structural resonances[9] in the whole range of $B_{ext}$, the data show distinct Hall plateau and Shubnikov-de Haas minima starting at 2.3 T for a filling factor of 6. Thus, there are four transverse modes in the wire when $B_{ext}$ < 2.3T and the magneto-depopulation from the forth mode to the third mode occurs at 2.3T.

The effective electron density in the wire is $3.6 \times 10^{15}$ m$^{-2}$, corresponding to a Fermi level of 30 meV. The estimated longitudinal wavelengths for the first three transverse modes are 44nm, 50nm and 66nm, smaller than the contact size that is about 80% of the ring diameter, determined from AFM images. As a result of the collimation effect, these three modes will not enter the ring until $B_{ext}$ is as high as 0.9T due to the magnetic focusing effect. While all devices show AB oscillations at low $B_{ext}$, rings with $R$ = 250nm and 350nm exhibit visible double frequency (*h/2e*) component in the raw conductance data.

Here we will focus on the device that shows double frequency around zero $B_{ext}$. Figure 2 (a) displays the AB interference for a 250nm ring at 1.9K with the background magnetoresistance subtracted. There are two distinct features of ΔR: (1) the unambiguous *h/2e* oscillations around zero $B_{ext}$; and (2) the quantum beating pattern with five visible transitions to the fundamental frequency *h/e*, where noticeable nodes are indicated by arrows in Fig. 2 (a). In the following, we show that these features in ΔR are experimental signatures of spin quantum phase. More specifically, they are the result of a superposition of two independent interference patterns due to the two orthogonal spin chiral states of the forth transverse mode.

From Eq. (1), it is clear that the observed double frequency in the vicinity of zero magnetic field, i.e. $w_Z \sim 0$, is a result of two conditions. (1) The existence of two chiral spin states, i.e., $w_{SO} \neq 0$. (2) Simultaneously, for this particular ring, an accidental coincidence: $2w_{SO}/\Omega \approx n+1/2$, where $n = 2,3,4, \ldots$ In other words, the spin dynamical phases of the two chiral states are out of phase after a single passage. In other rings that we have studied, the double-frequency signal manifests itself at different magnetic fields, determined by the ratio $w_{SO}/\Omega$ in those samples (see, e.g., Fig. 2(b)). The prominent quantum beating in the vicinity of zero magnetic fields constitutes a direct observation of the so-called "zero magnetic field spin splitting." This contrasts with the earlier measurements via extrapolation of Shubnikov-de Haas oscillations data in high magnetic field to zero field.[21]

We see that $\Delta G$ displays $h/2e$ oscillations when $\Delta j = j_+ - j_- = \pi, 3\pi, 5\pi\ldots$ (i.e., out-of-phase), and a fundamental $h/e$ period when $\Delta j = 2\pi, 4\pi, 6\pi\ldots$ (i.e., in-phase). In addition, it is the relative amplitudes of $W$, $w_Z$, and $w_{SO}$ that determine how fast $\Delta j$ experiences a $\pi$ change with $B_{ext}$. Assuming $g^* = -12$ for an InAs wire,[22] we find $\hbar w_Z = 0.35$ meV/T. Therefore, in order to have two chiral spin states experience the in-phase beating a few times within 1T, we need $\hbar W \ll 0.35$meV. The simulated $DG$ describes the data best when $\hbar W = 0.069$ meV and $2w_{SO}/W = 6.5$. Following Eqs. (1) and (2), the calculated $\Delta j$ and $DG$ are plotted respectively in Fig. 3 (a) and (b), where we assume $L_f = 3$ μm. It is clear that $DG$ displays a full $h/e$ period when $\Delta j$ is in-phase at fields marked by dots in Fig. 3 (a). In terms of the in-phase positions, the agreement between the data and the calculation is excellent. The parameters $W$ and $w_{SO}$ used in the simulation are very reasonable for this experimental configuration. The wave vector $k_0$ obtained from $W$, assuming $m^* = 0.033m_0$ is $7.5 \times 10^6$ m$^{-1}$. As a result, the kinetic energy is ~0.1meV. This small energy is consistent with the picture that the conductance oscillation beating comes solely from electrons in the fourth transverse mode. Moreover, the estimated effective $B_{in}$ and $\alpha$ are also reasonable: 0.64T and $3.0\times10^{-11}$ eVm, respectively.

To elucidate further the role of the spin Berry's phase, we attempted to model our data taking into account only the dynamical and the AB phases, i.e., excluding the Berry phase from Eq. (1). In order to have $\Delta\boldsymbol{j}$ in-phase beating occur around the same fields as in the data, we had to use $\hbar\boldsymbol{W} = 0.05$ meV and $2\boldsymbol{w}_{SO}/\boldsymbol{W} = 22.5$, even though the corresponding $\alpha$ becomes as large as $1.0\times10^{-10}$ eVm. The calculated $\Delta\boldsymbol{j}$ and $\boldsymbol{D}G$ are plotted in Fig. 3 (a) as dashed curve and in Fig. 3 (c), respectively. Although the $\boldsymbol{D}G$'s in Fig. 3 (b) and (c) look similar, there are two major distinctions in addition to the less realistic $\alpha$ for the curve obtained excluding Berry's phase. First, the in-phase positions are different: the first in-phase beating comes at a $B_{ext}$ lower than the observed, while the rest of such occurrences are at much higher fields, marked as diamonds in Fig. 3 (a). Second, $\boldsymbol{D}G$ in Fig. 3 (b) has an additional beating around $B_{ext} = 0.1$T, corresponding to the minimum of $\Delta\boldsymbol{j}$ depicted in the inset of Fig. 3 (a). Such a turn-around feature in $\Delta\boldsymbol{j}$ is due to the different dependence of the dynamical and Berry phases on $B_{ext}$. It can result in a pronounced signature in the oscillation data, as shown in Fig. 2(b) around 0.05T. Such characteristic in the raw data provides a direct, unambiguous evidence of the spin Berry's phase. We conclude that it is primarily the Berry phase that determines the behaviour of $\Delta\boldsymbol{j}$ and $\boldsymbol{D}G$ at low $B_{ext}$ and shifts the first appearance of the in-phase beating to higher fields. It is noteworthy that InAs, where the spin chiral states are primarily due to asymmetry of confinement[13] rather than due to the lowered symmetry in the host crystal,[23] is an ideal system for observing Berry's phase. If the corresponding two SO terms are comparable, the area enclosed by $\vec{B}_{eff}$ in parameter space for electron circumnavigating the ring is significantly reduced, and if these terms have equal magnitude, no area is enclosed and Berry's phase vanishes.

It is a popular practice to perform Fourier transform (FT) in search of Berry's phase, where a splitting or a sideband in the FT power spectrum are regarded as the evidence.[5,6,7] However, we show here that FT spectra exhibit a complicated dependence on the range of $B_{ext}$ taken to perform FT, and cannot be used as a sole evidence of the Berry phase. Figure 4 displays the evolution of FT spectra for (a) the experimental data shown in Fig. 2 (a), (b) the curve in Fig. 3 (b), and (c) the curve in Fig. 3 (c). As opposed to Fig. 4(c), Fig. 4(b) replicates well the experimental data. Nevertheless, the characters of the FT spectrum, such as the presence of a splitting or a sideband, can be modelled *irrespective* of the presence or absence of the spin Berry's phase. It is important to realize that the FT of Eq. (2) is significantly affected by $\Delta \boldsymbol{j}$ at $B_{ext} = 0$. In particular, if we arbitrarily add a constant phase to $\Delta \boldsymbol{j}$, a central peak rather than a dip appears in the FT spectrum even if the rest of the parameters remain the same. Therefore, the effectiveness of the FT power spectrum is limited. We emphasise that only the raw conductance data can bring a conclusive evidence for the spin Berry phase.

In conclusion, we have shown that two distinctive attributes, the singly-connected ring configuration and the collimating contact, provide a remarkable experimental setting for the measurement of the full spin quantum phase, including Berry's phase and the dynamical phase. We have demonstrated that the observed double-frequency features in AB oscillations are the superposition of conductance signals originated from two chiral spin states passing through the ring twice. The interplay of the Berry's and dynamical phases manifests themselves in transitions of AB conductance between double and single frequency oscillations. By comparing data with simulations, we have shown that the spin Berry's phase has a profound impact on the AB oscillations and on their Fourier power spectrum for $B_{ext} < B_{in}$. The observations of quantum beating and double-frequency oscillations indicate a long spin coherent length, more than 3 μm at a relatively high temperature of 1.9K. A ring with two spin chiral states is an interesting example of quantum two-level systems, which are currently the focus of attention as the building blocks of quantum computers. This system can also generate other prospects for future quantum technologies. Our work shows that interference signals in nanostructures can manifest themselves in conductance beating due to the long spin coherence length and can be modified by creating contacts that filter electron modes. With the on-going effort to create electrical gates to these devices, we anticipate further achievements in controlling and manipulating chiral spin states.

Acknowledgements: We thank T. L. Reinecke and B. V. Shanabrook for critical reading of the manuscript and W. J. Moore for discussion on Fourier transform. The work is supported in part by the ONR/NNI, LPS/NSA, ARDA, and DARPA.

**Figure Captions**

Fig. 1 Trajectories of electrons in real space and of their spins in parameter space for (a) a singly connected ring and (b) a doubly connected ring.

Fig. 2 (a) The experimental quantum beating pattern for a singly-connected InAs ring with a radius of 250nm at 1.9K. The arrows indicate the in-phase nodes for two spin chiral states. The right inset is a 2μm×2μm atomic force microscope image of an InAs ring. The left inset shows the magnetoresistance of an InAs ring at 2K. (b) The measured beating pattern for a 350nm ring where the additional beating feature around 0.05T is marked with dots.

Fig. 3 (a) The solid curve shows the simulated full spin phase for the data shown in Fig. 2 (a). The fitting parameters are $\hbar W = 0.069$ meV and $2w_{SO}/W = 6.5$. The dashed curve represents the simulation excluding Berry's phase. The fitting parameters are $\hbar W = 0.05$ meV and $2w_{SO}/W = 22.5$. The dots and diamonds mark the in-phase nodes. The curves have been offset by an integer for clarity. The inset amplifies the difference of these two simulations at low $B_{ext}$. (b) The simulated conductance oscillations taking into account the full spin phase, where the feature corresponding to the minimum of Δφ is marked with dots. (c) The simulation excluding Berry's phase, see text for discussion.

Fig. 4 The evolution of the Fourier transform spectra taken using different ranges of magnetic field, $\Delta B_{ext}$, for (a) the experimental data shown in Fig. 2(a), (b) the simulation including the full spin phase, and (c) the simulation excluding Berry's phase.

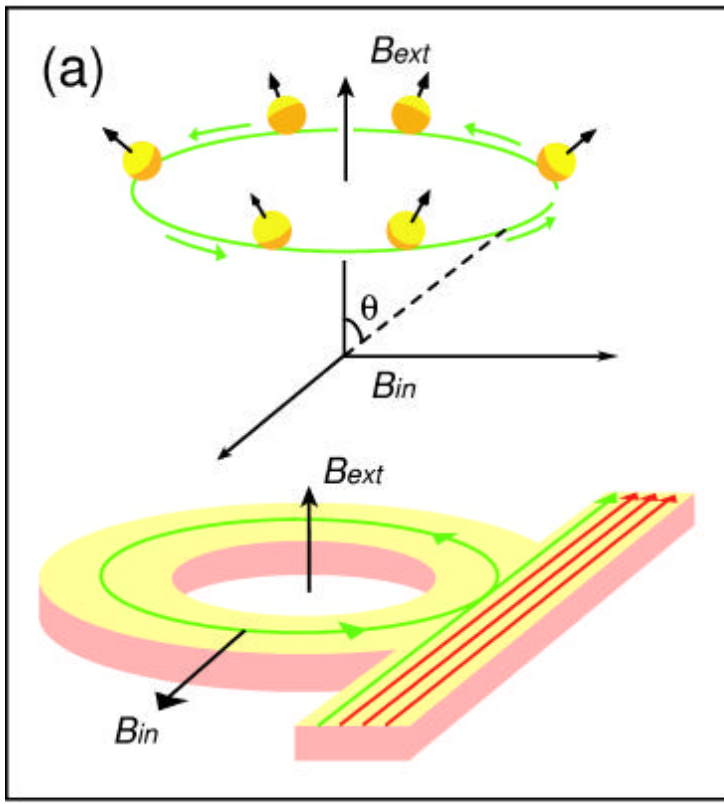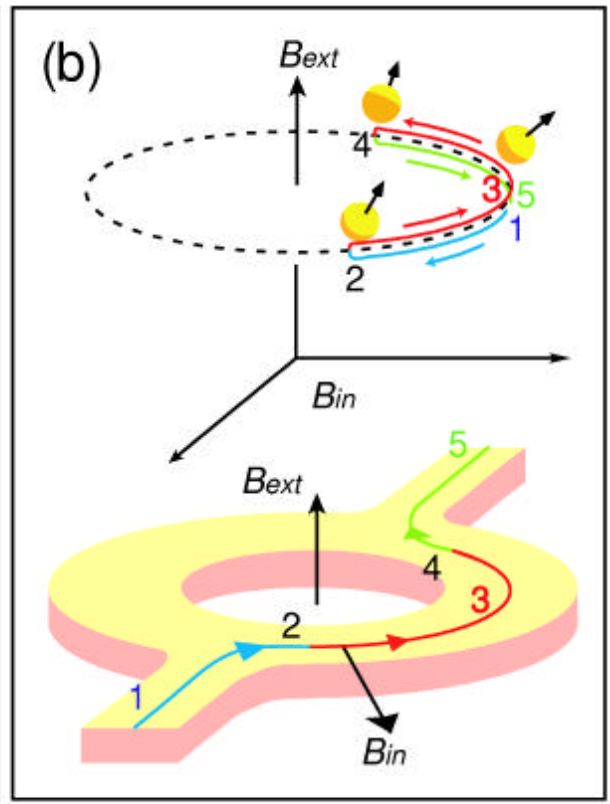

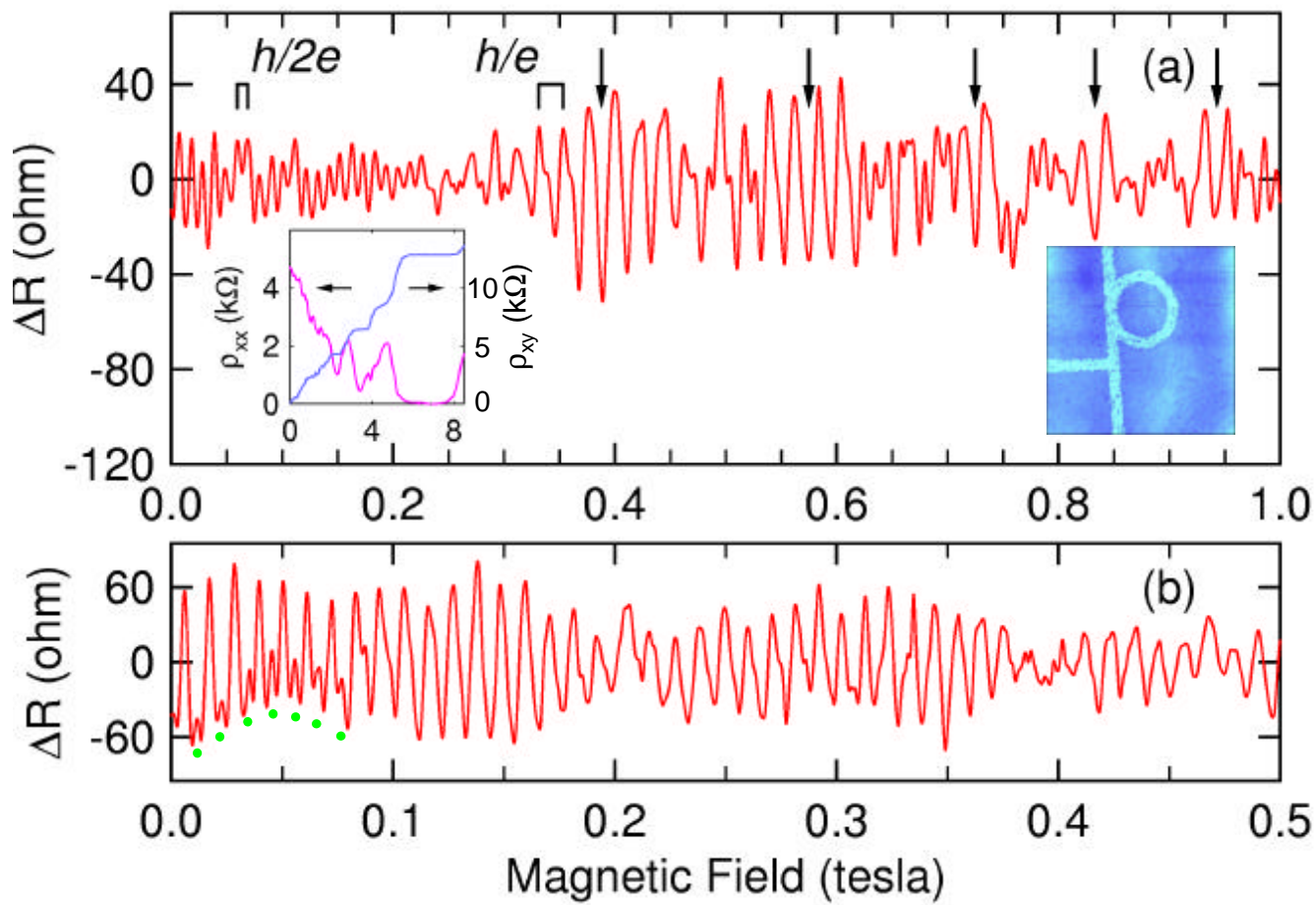

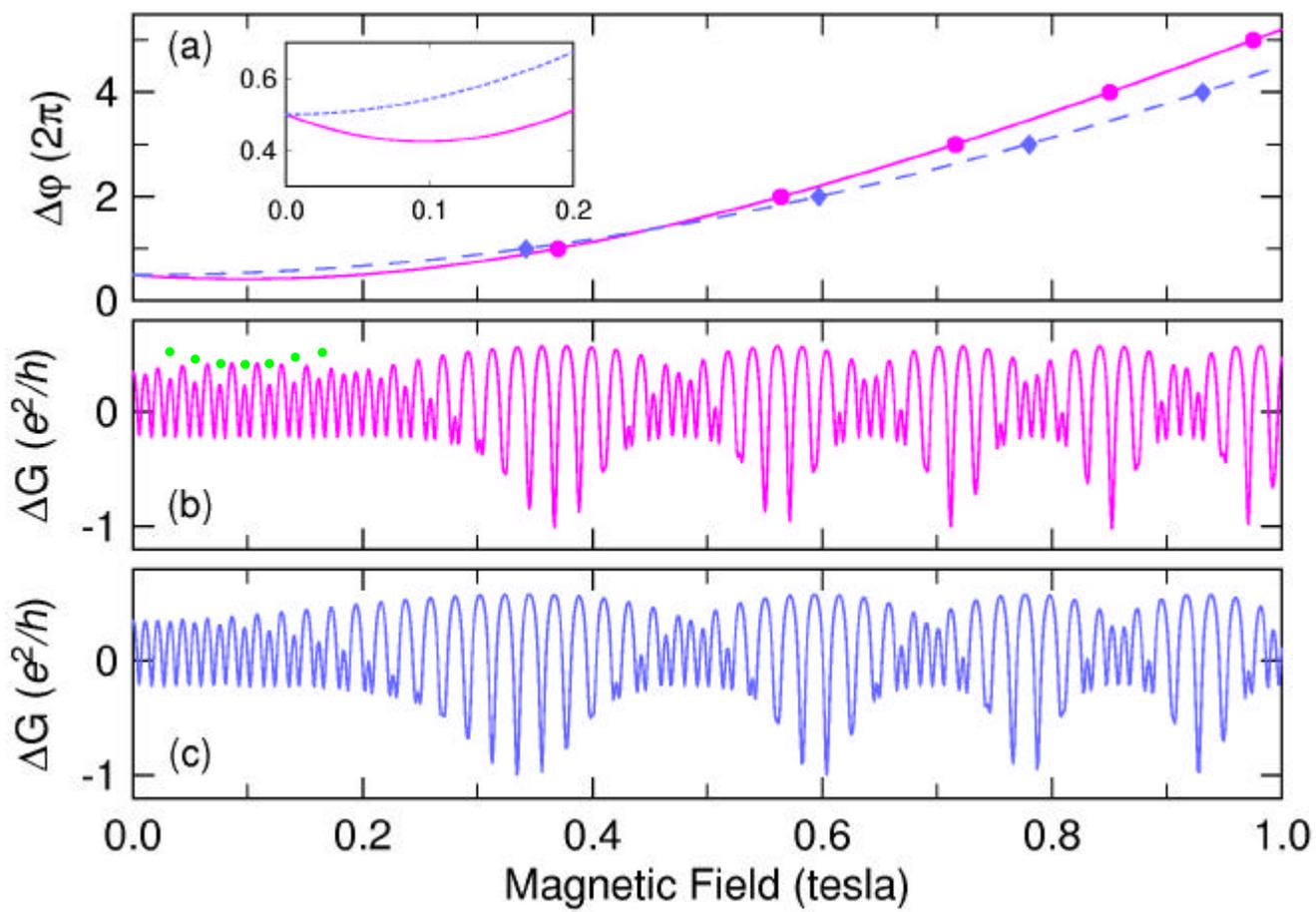

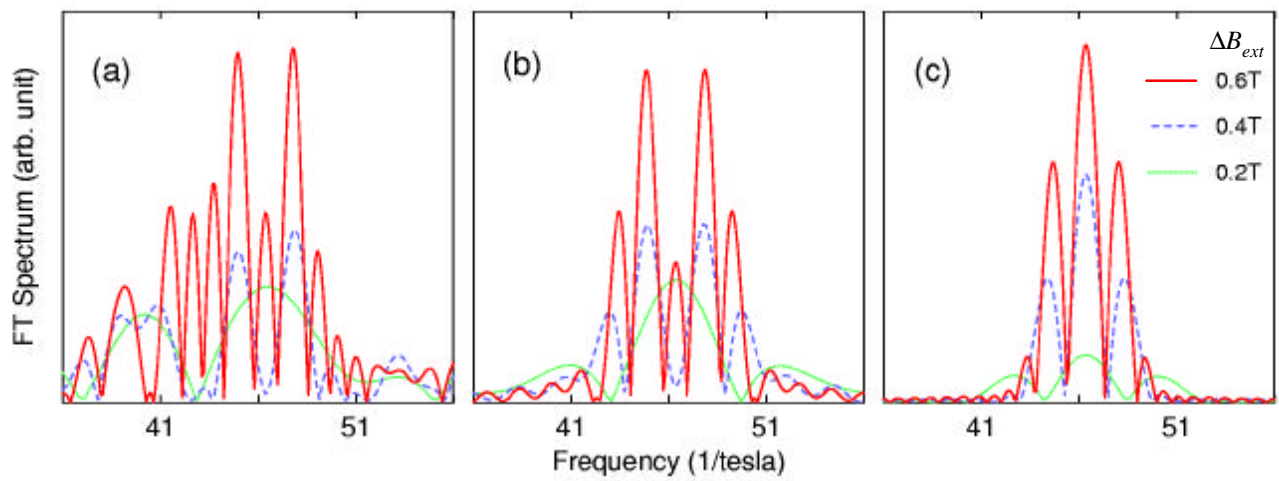